\documentclass[twocolumn]{article}
\usepackage[]{graphicx}  
\usepackage[a4paper, margin=0.8in]{geometry}

\usepackage{xcolor} % for pdf, bitmapped graphics files

\newcommand{\aff}[1]{$^{\mathrm{#1}}$}

\title{\LARGE \bf
Impulse Response Characterization of a Commercial Multimode Fiber using Superconducting Nanowire Single-Photon Detectors}

\author{
Yuanhang Zhang\aff{1,2}, Nicolas K. Fontaine\aff{1}, Mikael Mazur\aff{1}, Haoshuo Chen\aff{1}, Roland Ryf\aff{1},\\ Guifang Li\aff{2} and  Andrea Blanco-Redondo\aff{1}}

\vspace{3mm}
\date{%
    \small{
    \aff{1}Nokia Bell Labs, 600 Mountain Ave., New Providence, NJ 07974, USA\\
    \aff{2}CREOL, the College of Optics and Photonics, University of Central Florida, Orlando, FL 32816, USA.\\
    \textcolor{blue}{yuanhangzhang@knights.ucf.edu}} \\
    \today
}

\begin{document}
\maketitle

\thispagestyle{empty}
\pagestyle{empty}

%%%%%%%%%%%%%%%%%%%%%%%%%%%%%%%%%%%%%%%%%%%%%%%%%%%%%%%%%%%%%%%%%%%%%%%%%%%%%%%%
\begin{abstract}
Time-of-flight measurements are key to study distributed mode coupling and differential mode group delay (DMDG) in multimode fibers (MMFs). However, current approaches using regular photodetectors with limited sensitivity preclude the detection of weak modal interactions in such fibers masking interesting physical effects. Here we demonstrate the use of high-sensitivity superconducting nanowire single-photon detectors (SNSPDs) to measure the mode transfer matrix of a commercial graded-index multimode fiber. Two high performance 45-mode multi-plane light conversion (MPLC) devices served as the mode multiplexer/demultiplexer. Distributed mode coupling and DMDG among different mode groups are accurately quantified from the impulse response measurement. We also observed cladding modes of the MMF as a pedestal of the pulse in the measurement. This work paves way for applications such as quantum communications using many spatial modes of the fiber. 

\end{abstract}

%%%%%%%%%%%%%%%%%%%%%%%%%%%%%%%%%%%%%%%%%%%%%%%%%%%%%%%%%%%%%%%%%%%%%%%%%%%%%%%%
% \begin{keywords}
% Multi-plane light conversion (MPLC), Transfer matrix, Differential mode group delay (DMGD), Distributed mode coupling, Impulse response, Superconductive nanowire single-photon detector (SNSPD)
% \end{keywords}

% \fi
% \IEEEpeerreviewmaketitle

\section{Introduction}

Mode-division multiplexing (MDM) in multimode fibers (MMFs) as a candidate for next generation high-capacity optical transmission systems, has attracted great research interest in the last decade \cite{winzer2018fiber,li2014space}. The mode dynamics in MMFs are complex: spatial modes in these fibers couple with each other due to perturbations from the environment, imperfect geometry or refractive index fluctuations, and optical pulses propagating in MMFs suffer from modal dispersion, with each mode traveling at a different group velocity. The arrival time difference of two modes in a fiber is defined as differential mode group delay (DMGD) and it has been extensively studied in optical communications as it directly determines the number of equalizers taps necessary to compensate for the mode coupling \cite{arik2012effect}. Measuring the intensity impulse response is an important method to study DMGD and mode coupling.  
\begin{figure*}[!t]
    \centering
    \includegraphics[width = 5.5in]{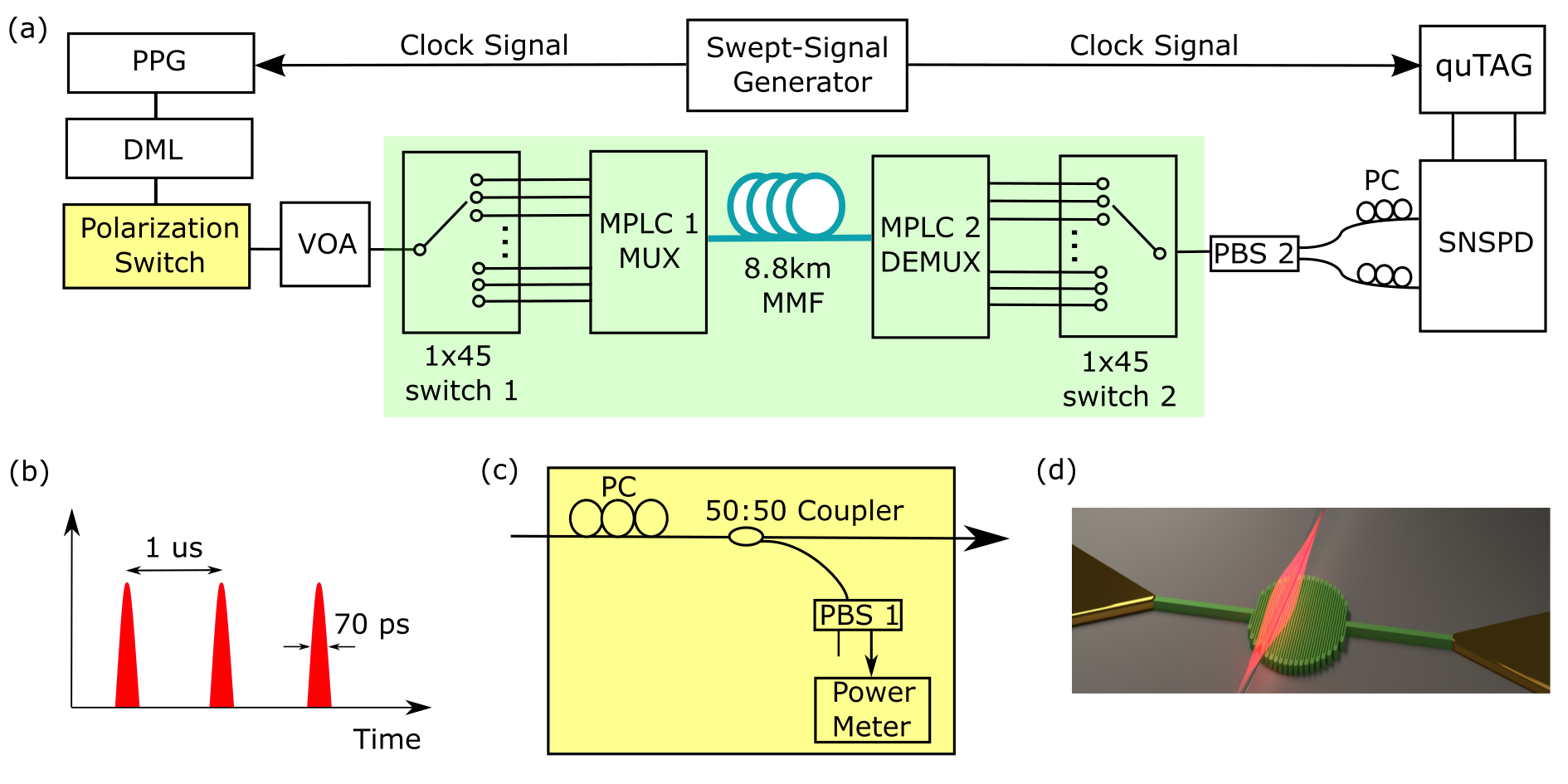}
    % where an .eps filename suffix will be assumed under latex, and a .pdf suffix will be assumed for pdflatex; or what has been declared
    % via \DeclareGraphicsExtensions.
    \caption{(a) Experimental setup for the transfer matrix measurement. PPG: pulse pattern generator; DML: directly modulated laser; VOA: variable optical attenuator; MPLC: multi-plane light converter; MUX/DEMUX: multiplexer/demultiplexer; PBS: polarization beam splitter; PC: polarization controller; SNSPD: superconductive nanowire single-photon detector. (b) Output periodic pulses of the DML. (c) Components used to make the polarization switch. (d) Artist impression of a SNSPD, with photons impinging on a superconductive meandering nanowire (Courtesy of Single Quantum).}
    \label{fig_setup}
\end{figure*}
%% ======================================================

Current methods to measure the impulse response of MMFs --- including time-of-flight techniques \cite{painchaud1992time,cheng2012time}, swept-wavelength interferometry (SWI) \cite{vanwiggeren2003swept}, optical time-domain reflectometry (OTDR) \cite{nakazawa2014measurement}, and digital signal processing (DSP) \cite{ryf2012impulse} --- fail to measure long fiber spans and very weak inter-modal interactions due to the limited sensitivity of regular photodetectors. The capability of detecting light at the single-photon level has enabled breakthroughs in many fields, such as reconfigurable photonics \cite{gyger2021reconfigurable}, light detection and ranging (LiDAR) \cite{shangguan2017dual}, and optical quantum information applications \cite{hadfield2009single} including quantum computing \cite{zhong2020quantum,zhong2021phase} and quantum key distribution \cite{takesue2007quantum}. Recently, using ultrasensitive single-photon detectors to study fiber modes has attracted some attention. In 2019, Johnson \textit{et al.} measured the temporal evolution of orbital angular momentum (OAM) modes at telecommunication wavelengths using a single-pixel camera  with a single-photon avalanche diode (SPAD) detector and a digital micro-mirror device (DMD) \cite{johnson2019light}. In 2020, Chandrasekharan \textit{et al.} measured the group velocity dispersion, DMGD, and the effective refractive index difference of different spatial modes in a 6-mode fiber using a two-dimensional (2D) SPAD array, over a broad bandwidth (500 nm -- 610 nm) in the visible regime \cite{chandrasekharan2020observing}. However, 2D SPAD arrays are not available at telecommunications wavelengths because of the high cost of fabricating pixelated arrays. Further superconducting nanowire single-photon detectors (SNSPDs) show significantly better performance at these wavelengths. At 1550nm,  commercially available SPADs have a timing jitter of $\sim$150 ps, quantum efficiencies of $\sim$20-30 \%, and dark count rates of $\sim$50 Hz \cite{SPADwebsite}. In contrast, the commercial SNSPD (Single Quantum Eos) in our group has a timing jitter of $\sim$10 ps, a quantum efficiency of $\sim$85\%, and dark count rates of $>$10 Hz. Notably, SNSPDs with sub-3 ps temporal resolution have been demonstrated \cite{korzh2020demonstration}.  In 2021, Mazur \textit{et al.} did a proof-of-principle impulse response measurement of a few-mode fiber using the SNSPD and a 3-mode photonic lantern as the mode demultiplexer \cite{mazur2021impulse}. 

In this paper, we measure the impulse response of the dual-polarization 45$\times$45 transfer matrix of a commercial OM3 multimode fiber. With the high sensitivity of SNSPDs, we build the histogram of the impulse response at a single-photon level, enabling us to study very weak modal interactions. The very low dark count rate used ($\sim$200 Hz) determines a noise floor of -136 dBm, giving us the capability to observe distributed mode crosstalk and cladding modes directly at the foot of the pulse. Using the high timing accuracy of $\sim$ 10 ps, the DMGD can be accurately measured. Two high performance 45-mode multiplexer/demultiplexer based on MPLC technology and two optical MEMS (micro-electromechanical system) switches are used to excite and receive one specific spatial mode of the MMF at every single step. 

This paper is structured into four sections. Section II describes the experimental setup and some critical components used in the measurement. Section III analyzes the results of two cases: the back-to-back (BTB) transmission and transmission with the MMF inserted. DMDG and distributed crosstalk plateau among different mode groups are accurately quantified. Finally, section IV concludes the paper. 

\section{Experimental Setup}

%\IEEEpubidadjcol
% ================ figure 2 ====================
\begin{figure*}[!t]
\centering
\includegraphics[width = 7in]{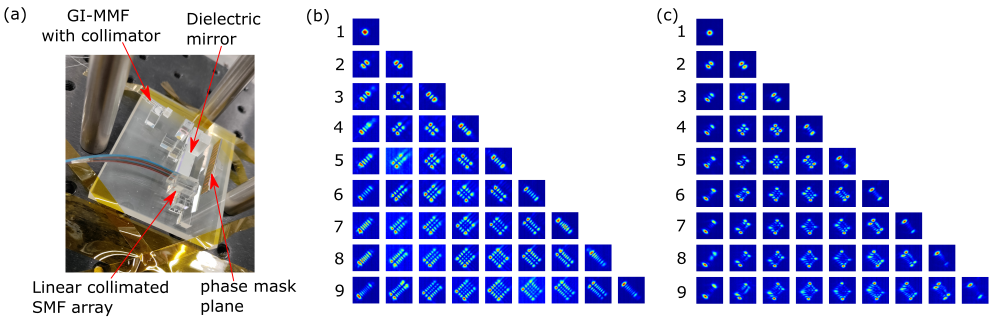}
% where an .eps filename suffix will be assumed under latex, and a .pdf suffix will be assumed for pdflatex; or what has been declared
% via \DeclareGraphicsExtensions.
\caption{(a) Photograph of one of the 45-mode MPLC devices used in the experiment The multi-mode output end of this device has a 3-m long GI-MMF with a collimator.  Free-space output modes of (b) MPLC 1 and (c) MPLC 2 captured by an InGaAs camera. Modes of 2nd MPLC are not as clear as the 1st MPLC due to imperfect packaging. 45 Hermite-Gaussian modes are grouped into 9 mode groups. Modes within the same group are degenerate and tend to couple strongly during propagation in a fiber. }
\label{fig_MPLC}
\end{figure*}
% ================================================
The experimental setup is outlined in Fig. 1. A swept-signal generator (Agilent 83650B) provided a sinusoidal clock signal (8000 MHz, 125 ps) for the pulse pattern generator (PPG, Anritsu MP1763B). The PPG divided the clock signal by 8000 (1MHz, 1$\mu$s) and generated square pulses to drive a directly modulated laser (DML). This DML is a distributed feedback (DFB) laser with a measured wavelength of 1545 nm. The swept-signal generator also provided an external clock signal for the quTAG (qutools company) for synchronization. The DML has a pulse repetition rate of 1 MHz, and each pulse has a measured full width at half maximum (FWHM) of 70 ps. We chose a repetition rate of 1 MHz as the detectors have peak efficiency ($\sim$85\%) at this rate, higher rates are measurable at the expense of lower detection efficiency. The average output power of the DML was measured to be -36.0 dBm.  To fully characterize the dual-polarization 45$\times$45 transfer matrix, a manually controlled polarization switch, shown in Fig. 1(c), was used in front of the mode multiplexer, and a fiber-based polarization beam splitter (PBS) was used after the mode demultiplexer to realize a polarization-diverse measurement. The polarization switch, as shown in Fig. 1(c), consists of a polarization controller (PC), a 50:50 fiber coupler, a fiber-based PBS (Thorlabs, PBC1550SM-APC) and a power meter (HP 8153A). A pair of optical switches (DiCon MEMS 64-channel model) and MPLCs constitute the function of mode-diverse measurement (green background in Fig. 1(a)). Outputs from the second PBS were connected to two independent channels of the SNSPD system. Since the meandering nanowire inside the SNSPD is polarization sensitive, two PCs were added between the second PBS and the SNSPD. Any path length mismatch (either in fiber or RF cable) induced a relative time shift for pulses in different channels, which was corrected during the post-processing. 

To launch the $x$-polarization state into the system with the polarization switch shown in Fig. 1(c), we first adjusted the PC manually until reading the maximum  power (-39.6 dBm) on the power meter. Then we scanned all 45 channels of switch 1 and switch 2 sequentially, with the received $x$ and $y$ polarization states recorded simultaneously under $x$-polarization launch. Following, we adjusted the PC until reading the minimum power (-70.0 dBm), to launch $y$-polarization state and repeated the above mode-diverse measurement. In this way, we make sure $x$- and $y$- polarization launch states are orthogonal. Those two optical switches and the quTAG were controlled by one single computer to record data. For every cell of the 45$\times$45 matrix, 30~s of raw data was saved to accumulate enough data points to plot the histogram. Measuring a complete dual-polarization transfer matrix needed roughly 34 hours. We measured the 45$\times$45 transfer matrix in a back-to-back (BTB) experiment first (splicing the pigtail fibers of two MPLCs directly \cite{fontaine2018packaged}). After taking data of the entire transfer matrix of the BTB, we inserted a spool of commercial MMF and repeated the measurement. The VOA helped to attenuate light to single-photon level. It had an attenuation of 33 dB for the BTB measurement and was adjusted to 30 dB after inserting the MMF. In this way, the SNSPD reads a similar photon count rate for BTB and MMF measurement.

In the BTB when both switches were switched to channel 1 ($HG_{00}$ mode to $HG_{00}$ mode coupling), the total power reaching the SNSPD was estimated to be -95 dBm. Since such a low power cannot be measured directly with a normal photo detector, we set the VOA to 0 dB attenuation first, and measured the power before the second PBS being -62 dBm. The total insertion loss of these two switches and MPLCs was measured to be 18 dB. Other losses in the link are from the 50:50 fiber coupler, connectors, and some single mode fiber (SMF) connecting components between two labs.

% The SNSPDs are cooled down by helium to 2.6 K inside a closed-cycle cryostat, providing a low temperature environment for the superconducting nanowires. The superconducting nanowires are biased with a direct current close to the critical current, above which the superconductivity of the nanowire breaks down. Part of the detector transits from a superconducting state to a resistive state upon absorption of a photon. This transition will give a fast electrical pulse that is amplified by the SNSPD driver. 
The SNSPD was cooled down by helium to 2.6 K and the bias current of each individual detector was set just below the critical current of the nanowires, leading to a dark count rate of about 200 counts/s, corresponding to a sensitivity of -136 dBm. The detector would transit from a superconducting state to a resistive state upon absorption of a single photon.  In our experiment, all lights in the lab were turned off when calibrating the dark count rate, and most of the setup including the MMF fiber spool and MPLCs were covered with aluminum foil to reduce the influence from surrounding light. Covering the setup is important to reach a low dark count rate because background light may couple into the fiber due to blackbody radiation at room temperature \cite{shibata2015ultimate}. 
%% ============ figure 3 ==========================
\begin{figure*}[!t]
    \centering
    \includegraphics[width = 4.6in]{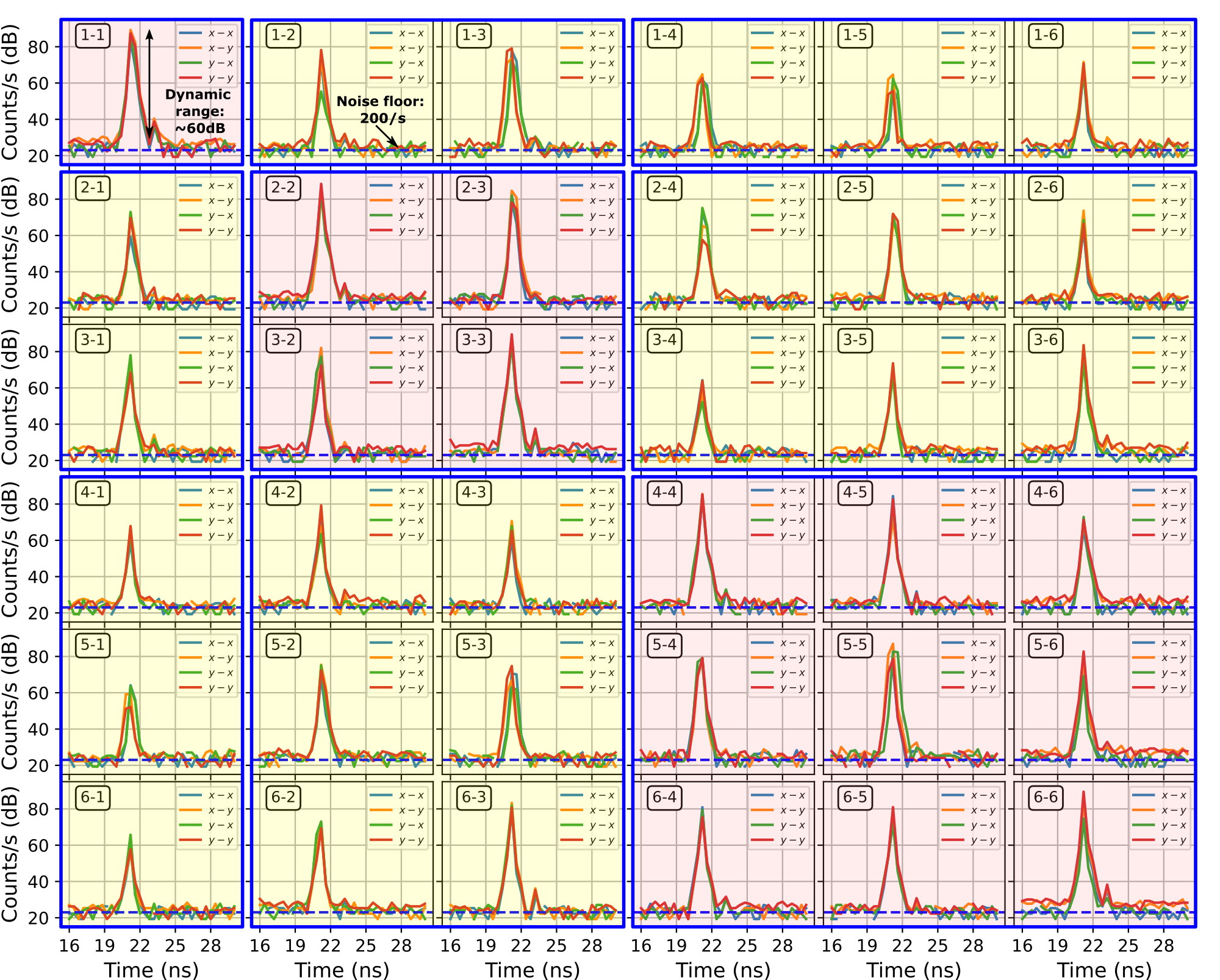}
    \caption{Histograms of photon arrival time of the BTB experiment. This figure shows the 6$\times$6 matrix of the up-left corner of the entire 45$\times$45 transfer matrix, with all polarizations presented in one subplot. Readers interested in the complete 45$\times$45 matrix can find it in the \textbf{supplementary figure 1}. The number (format 'A-B') in the round box in each subplot denotes the coupling from mode A to mode B. The blue dash line indicates the dark count rate of 200 counts/s (noise floor). Cells with the pink background color represent mode coupling of the same mode group while yellow color represent mode coupling among different groups}
    \label{fig_BTB_6x6}
\end{figure*}
%% ============== figure 4 ========================
\begin{figure*}[!ht]
    \centering
    \includegraphics[width = 4.6in]{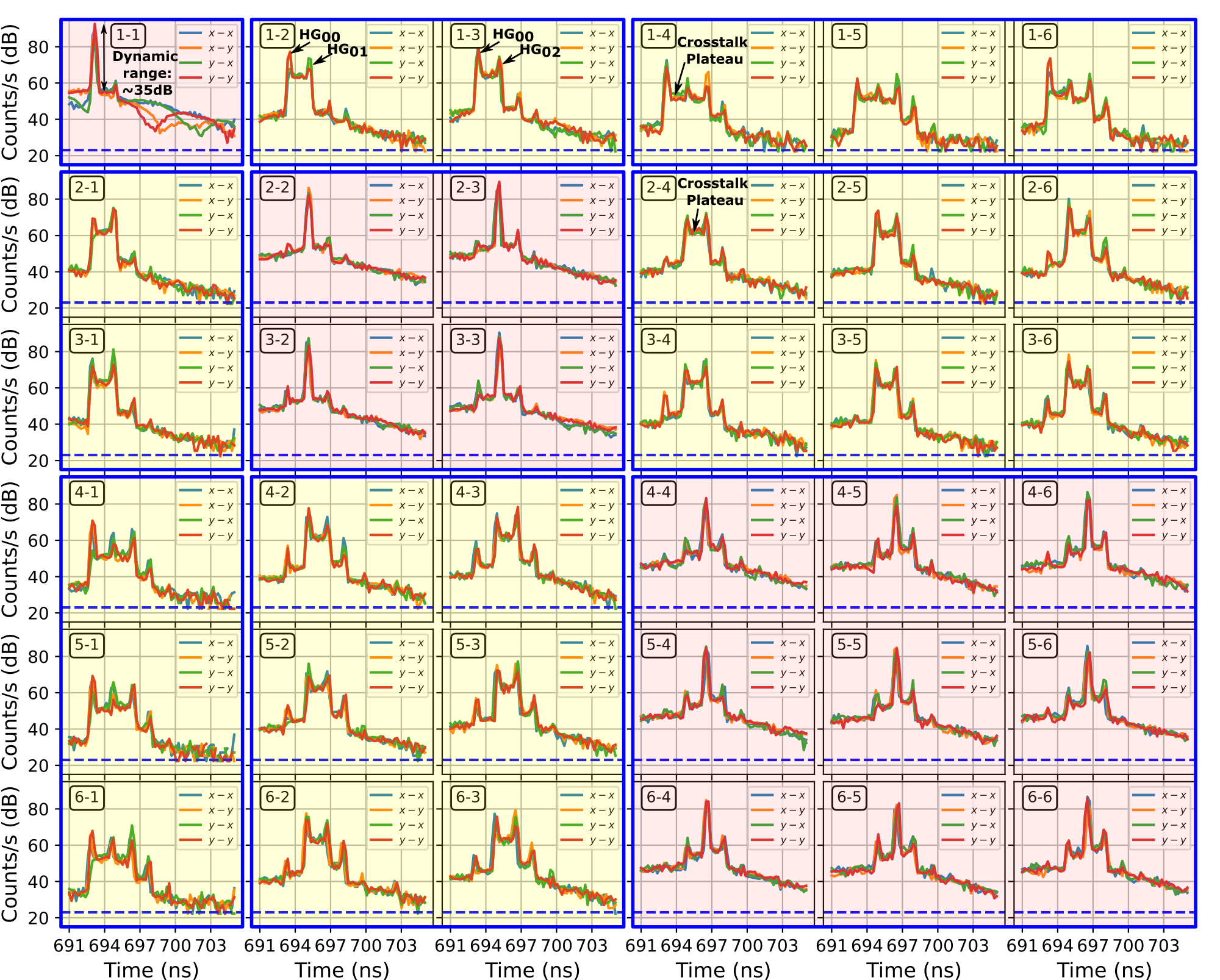}
    \caption{Histograms of photon arrival time with the MMF inserted. This figure shows the 6$\times$6 matrix of the up-left corner of the entire 45$\times$45 transfer matrix, with all polarizations presented in one subplot. Readers interested in the complete 45$\times$45 matrix can find it in the \textbf{supplementary figure 2}. The number (format 'A-B') in the round box in each subplot denotes the coupling from mode A to mode B. The blue dash line indicates the dark count rate of 200 counts/s (noise floor). Cells with the pink background color represent mode coupling of the same mode group while yellow color represent mode coupling among different groups.} 
    \label{fig_MMF_6x6}
\end{figure*}
%% ==================figure 5 ===============================
\begin{figure*}[!ht]
    \centering
    \includegraphics[width = 6.5in]{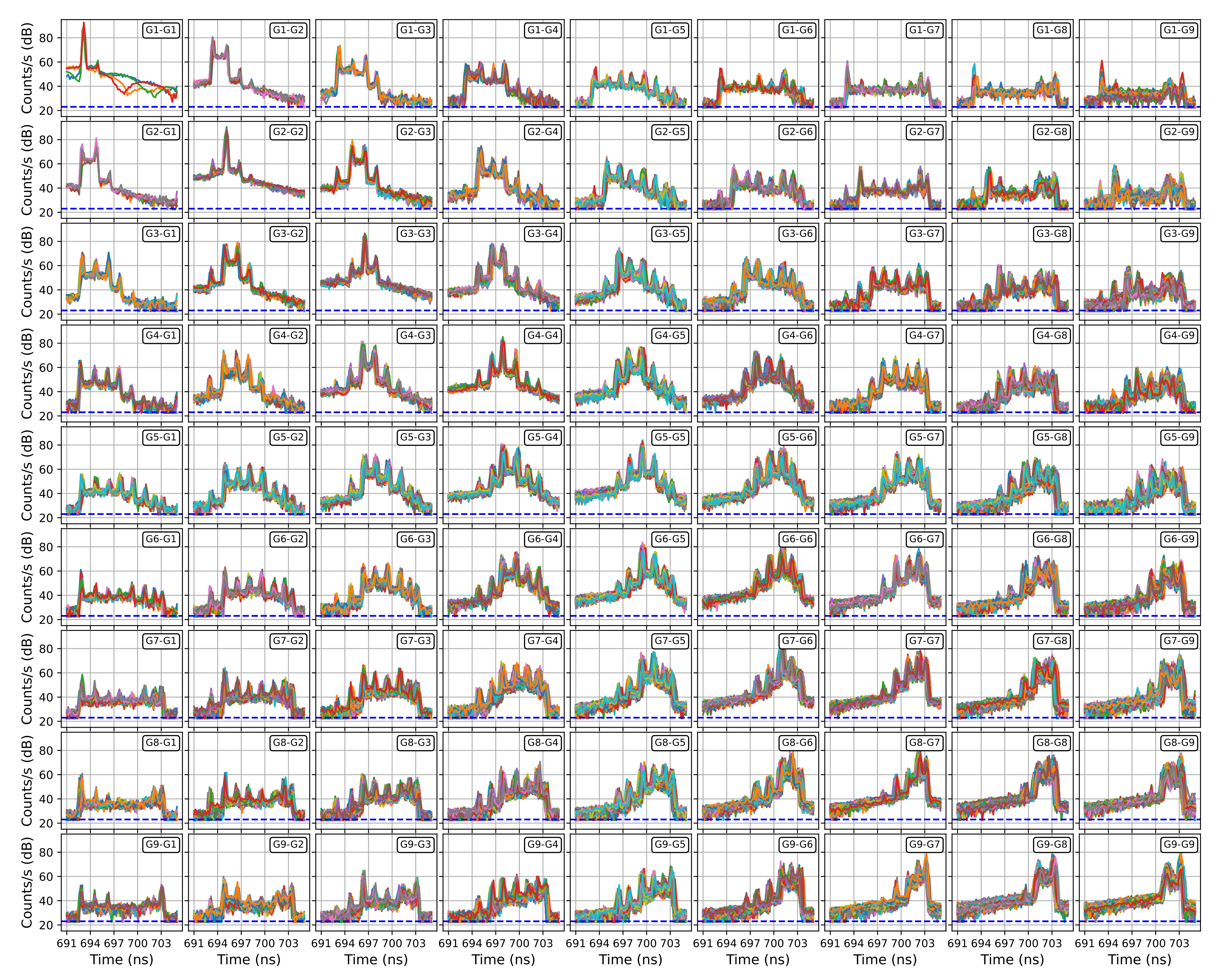}
    \caption{Histograms of mode group coupling. The entire 45$\times$45 matrix is grouped into a 9$\times$9 matrix. All polarizations and coupling pulses belong to the same mode group are plotted in one subplot. The blue dash line indicates the dark count rate of 200 counts/s (noise floor).}  
    \label{fig_MMF_9x9}
\end{figure*}

%% ===================================================
This SNSPD system has 4 independent channels, of which channel 3 and channel 4 were used in the experiment to register the received light with $x$ and $y$ polarizations separately. At 1550 nm, channel 3 and 4 has a system detection efficiency of 81\% and 85\%, respectively. The calibrated timing jitter of channels 3 and 4 are 19.0 ps and 13.2 ps FWHM, respectively. The measured relaxation time (full recovery time after each detection event) is about 100 ns, limiting the maximum count rate to roughly 10 MHz. The detection speed therefore, is much lower than the regular balanced photodetectors (BPDs) commonly used in coherent optical communications, which can have a bandwidth up to 100 GHz. The quTAG is a fast time-to-digital converter and time tagging device for time correlated single-photon counting (TCSPC). The quTAG has a measured single channel timing jitter of 5.94 ps RMS and 14.00 ps FWHM.

A spool of commercially available OM3 graded-index multimode fiber (GI-MMF, OFS LaserWave FLEX 300) was measured in the experiment. The fiber has a total length of 8851m, and it was believed to support 9 mode groups (45 spatial modes) at 1550 nm. In this experiment however, we found it supports 10 mode groups (55 spatial modes), and the last two mode groups behave differently from the first 8 mode groups, as shown in Fig. 7(b). The same fiber has been used in a mode-division multiplexed transmission experiment in 2014 by Ryf \textit{et al.} \cite{ryf2015mode}. According to \cite{ryf2015mode}, this OM3 MMF has a measured attenuation loss of 0.34 dB/km for the $LP_{01}$ and $LP_{11}$ modes at 1550 nm, and a chromatic dispersion parameter of $\sim$20-24 ps/nm/km. However, accurate measurements of the distributed crosstalk, DMGD, and the cladding modes of this kind of fiber remained elusive, as they were beyond the sensitivity and temporal resolution of previously used detectors.

We use two MPLCs as the mode multiplexer/demultiplexer. The 45-mode MPLC is a free space device consisting of a collimated linear SMF array, a phase mask plane, a dielectric mirror, and a collimated GI-MMF at the output, as shown in Fig. 2(a).  The device maps every input Gaussian spot from the SMF array into a specific Hermite-Gaussian (HG) mode through a unitary transformation \cite{fontaine2018packaged,fontaine2017design}. As a reciprocal device, this device can be used either as a mode multiplexer (input through SMF end) or a mode demultiplexer (input through GI-MMF end). The performance of the two MPLCs are qualitatively shown by imaging the free space output modes using an InGaAs camera, as shown in Fig. 2 (b) and (c). The HG modes in Fig. 2(c) are not as clear as those in Fig. 2 (b), indicating the 2nd MPLC that used as the mode demultiplexer has a compromised performance compared to the first one. This was due to an imperfect packaging after assembling, leading to a relatively large insertion loss (IL) and mode dependent loss (MDL). 45 HG modes are grouped into 9 mode groups, and labeled in an order as shown in TABLE I. Modes within the same group are degenerate (having nearly equal propagation constant, and thus similar phase velocity and group velocity) and tend to couple strongly during propagation in a fiber. Actually, the 3-m pigtail fibers at the output end of two MPLCs are already long enough to induce mode coupling within the same mode group \cite{fontaine2018packaged}.
% \begin{center}
% \begin{tabular}{ c c c c c c c c c c }
%  $HG_{00}$ \\ 
%  $HG_{01}$ & $HG_{10}$  \\  
%  $HG_{02}$ & $HG_{11}$ & $HG_{20}$ \\
%  $HG_{03}$ & $HG_{12}$ & $HG_{21}$ & $HG_{30}$ \\
%  $HG_{04}$ & $HG_{13}$ & $HG_{22}$ & $HG_{31}$ & $HG_{40}$ \\
%  $HG_{05}$ & $HG_{14}$ & $HG_{23}$ & $HG_{32}$ & $HG_{41}$ & $HG_{50}$ \\
%  $HG_{06}$ & $HG_{15}$ & $HG_{24}$ & $HG_{33}$ & $HG_{42}$ & $HG_{51}$ & $HG_{60}$ \\
%  $HG_{07}$ & $HG_{16}$ & $HG_{25}$ & $HG_{34}$ & $HG_{43}$ & $HG_{52}$ & $HG_{61}$ & $HG_{70}$ \\
%  $HG_{08}$ & $HG_{17}$ & $HG_{26}$ & $HG_{35}$ & $HG_{44}$ & $HG_{53}$ & $HG_{62}$ & $HG_{71}$ & $HG_{80}$ \\
% \end{tabular}
% \end{center}

\begin{table}[!ht]
% \renewcommand{\arraystretch}{1.3}
% if using array.sty, it might be a good idea to tweak the value of
% \extrarowheight as needed to properly center the text within the cells
\caption{45 HG modes supported by the MPLC}
\label{table_HGmodes}
% Some packages, such as MDW tools, offer better commands for making tables
% than the plain LaTeX2e tabular which is used here.
% \rule{0pt}{10pt}
{\scriptsize
\renewcommand{\arraystretch}{1.3}
\begin{tabular}{c|l}
\hline
\hline
 Group 1 & $HG_{00}$ \\ 
 Group 2 & $HG_{01}$  $HG_{10}$  \\  
 Group 3 & $HG_{02}$  $HG_{11}$  $HG_{20}$ \\
 Group 4 & $HG_{03}$  $HG_{12}$  $HG_{21}$  $HG_{30}$ \\
 Group 5 & $HG_{04}$  $HG_{13}$  $HG_{22}$  $HG_{31}$  $HG_{40}$ \\
 Group 6 & $HG_{05}$  $HG_{14}$  $HG_{23}$  $HG_{32}$  $HG_{41}$  $HG_{50}$ \\
 Group 7 & $HG_{06}$  $HG_{15}$  $HG_{24}$  $HG_{33}$  $HG_{42}$  $HG_{51}$  $HG_{60}$ \\
 Group 8 & $HG_{07}$  $HG_{16}$  $HG_{25}$  $HG_{34}$  $HG_{43}$  $HG_{52}$  $HG_{61}$  $HG_{70}$ \\
 Group 9 & $HG_{08}$  $HG_{17}$  $HG_{26}$  $HG_{35}$  $HG_{44}$  $HG_{53}$  $HG_{62}$  $HG_{71}$  $HG_{80}$ \\
\hline
\hline

\end{tabular}
}
\end{table}

%% ==============================================

\section{Results and Discussion}
30s of time-of-flight measurement data were saved per polarization per mode of the 45$\times$45 transfer matrix. The data was used to build a histogram covering 1 $\mu$s, i.e., the pulse period of the DML. A zoom-in time window around the pulse of the entire 45$\times$45 matrix are plotted in the \textbf{supplementary figure 1} for BTB and \textbf{supplementary figure 2} for MMF. 

%%=================== figure 6 =================
\begin{figure}[!t]
    \centering
    \includegraphics[width=3in]{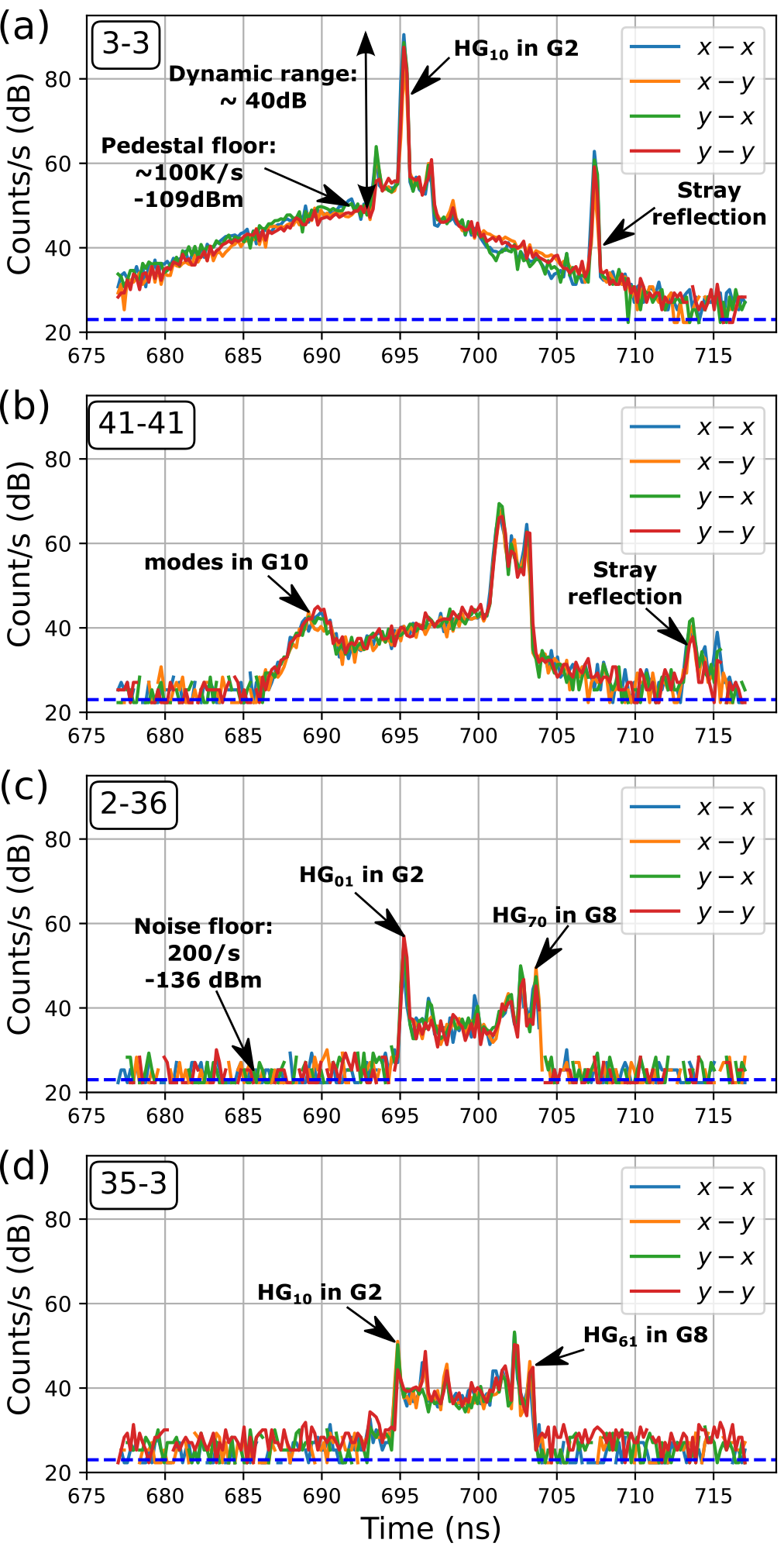}
     \caption{Selected elements from the entire 45$\times$45 transfer matrix with the MMF inserted: (a) 3-3 coupling, (b) 41-41 coupling, (c) 2-36 coupling and (d) 35-3 coupling. For some elements, e.g. the 2-36 at the off-diagonal region of the 45$\times$45 matrix, the pedestal is not obvious as in other elements. The blue dash line indicates the dark count rate of 200 counts/s.}
    \label{fig_selected_cells}
\end{figure}
%% =============================================
Fig. 3 shows the histograms of the 6$\times$6 matrix of the up-left corner of the entire 45$\times$45 transfer matrix of the BTB measurement, with all polarizations presented in one subplot. The peak appears at about 21 ns is the received pulse from the DML. Without a noticeable pedestal at the foot of the pulse, the dynamic range is as large as 60 dB. These subplots are grouped into different blocks, with cells of the pink background color represent mode coupling of the same mode group and yellow color represent mode coupling among different mode groups. 

Similar to Fig. 3, Fig. 4 shows the histograms of the 6$\times$6 matrix on the up-left corner of the entire 45$\times$45 transfer matrix with the MMF inserted. Distributed mode coupling and DMGD can be observed by the zoom-in view in the time window between 691 and 705 ns. Modes within the same group (e.g., $HG_{01}$ and $HG_{10}$) arrive at the same time due to degeneracy and thus have overlapped peaks in time. The crosstalk plateau between individual mode peaks is a clear sign of mode mixing. The plateau between the peaks is due to photons that experienced coupling from one mode to another mode somewhere along the fiber during propagation, with an arrival time between that of the photons that did not experience mode coupling. Same as Fig. 3, we use pink and yellow background color to represent mode coupling of the same mode group and among different mode groups, respectively. 

%% ================= figure 7 ========================
\begin{figure}[!t]
    \centering
    \includegraphics[width=3.3in]{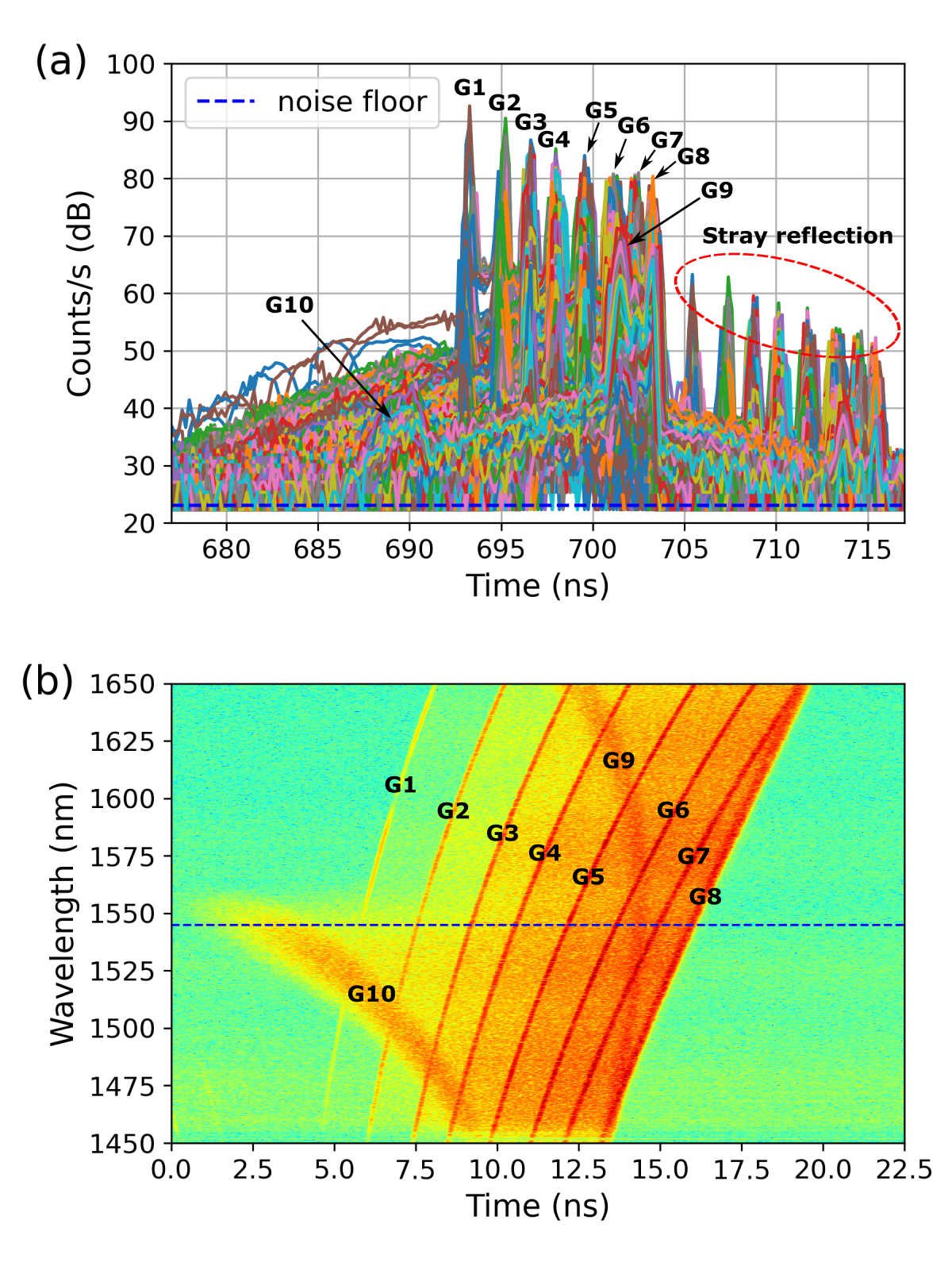}
    \caption{(a) Plotting all curves of the 45$\times$45 cells of the SNSPD data in one subplot. Only 8 mode group peaks are clearly distinguished. (b) Spectrogram of the same MMF measured using a swept-wavelength interferometer. The blue dash line indicates the laser wavelength (1545 nm) used in the SNSPD measurement.}
    \label{fig_SNSPD_OVNA}
\end{figure}
%% ===================================================
In Fig. 5, all polarizations and mode coupling curves within the same colored block in Fig. 4 are grouped in one subplot. In this way we group the original 45$\times$45 matrix into a 9$\times$9 matrix, considering degeneracy of modes in the same group. Fig. 6 shows some selected cells with an enlarged view. In the measurements with the MMF, the dynamic range is about 35$\sim$45 dB, limited by a pedestal at the foot of the pulse. As shown in Fig. 5 and Fig. 6, the value of the pedestal changes for different elements in the 45$\times$45 matrix. The general trend is that diagonal elements, which have larger count rates, also possess a higher pedestal level, such as in Fig. 6(a) and (b). Some off-diagonal elements, e.g., Fig. 6 (c) and (d), the pedestal is below the noise floor determined by the dark count rate ($\sim$200 counts/s, -136 dBm). Histograms of the BTB data in Fig. 3 lack the pedestal, indicating that the pedestal is not due to the limited extinction ratio of the DML \cite{mazur2021impulse}, but most likely to the cladding modes of the MMF. The silica cladding of the fiber can support hundreds of cladding modes and bending or other environment perturbations may violate the total internal reflection condition and cause mode energy to leave the core mode and couple to cladding modes~\cite{ivanov2006cladding}. The coupling back-and-forth between core modes and cladding modes during propagation and coupling among different coils of the fiber spool are responsible for the presence of cladding mode. For some cells (e.g. 41-41 in Fig. 6(b)) at the bottom-right corner of the 45\(\times\)45 MMF matrix (see \textbf{supplementary figure 2}), a "bump" appears at about 690 ns, 3 ns to the left of mode group 1. Further measurements confirm the "bump" is actually energy coupled to mode group 10, as discussed below in Fig. 7(b). 

Fig. 7(a) plots all curves of the 45$\times$45 cells of the MMF measurement together, with all curves well aligned in time. However, only 8 mode group peaks, instead of 9, are distinguishable in the plot. Note the 8 small peaks following the 8 main peaks are due to stray reflection in the system, as discussed in the supplementary part.  To investigate the reason we measured the same MMF with a swept-wavelength interferometer from 1450 nm to 1650 nm \cite{fontaine2013characterization}, shown as a spectrogram in Fig. 7(b). 10 bright lines represent 10 mode groups supported by the MMF in this wavelength range. The blue dash line at 1545 nm is the wavelength used in the SNSPD measurement. Clearly, mode group 9 (G9) lies in the crosstalk plateau between G6 and G7 at 1545 nm. This explains why only 8 mode groups peaks appear in Fig. 7(a). And note that mode group 10 (G10) arises as a "bump" about 3 ns to the left of mode group 1 (G1). The cut-off frequency of G10 is about 1555 nm. Lines corresponding to higher order mode groups are broader than lower order mode groups due to the existence of more vector modes with slightly different propagation constants in those mode groups \cite{fontaine2013characterization}. The slope of the lines shows chromatic dispersion, with modes in G9 and G10 having opposite dispersion sign than that those in groups G1 to G8 because they are less confined in the fiber core.

The DMGD values can be easily read out from the impulse response graph in Fig. 7(a) or the spectrogram in Fig. 7(b). The relative delay of G2, G3, G4, G5, G6, G7 and G8 to G1 are 1.97 ns, 3.35 ns, 4.73 ns, 6.31 ns, 7.88 ns, 9.07 ns and 10.06 ns, respectively, close to the DMGD numbers reported in a previous publication using the same fiber \cite{ryf2015mode}.

%% ===================== figure 8 =======================
\begin{figure}[!ht]
    \centering
    \includegraphics[width=2.7in]{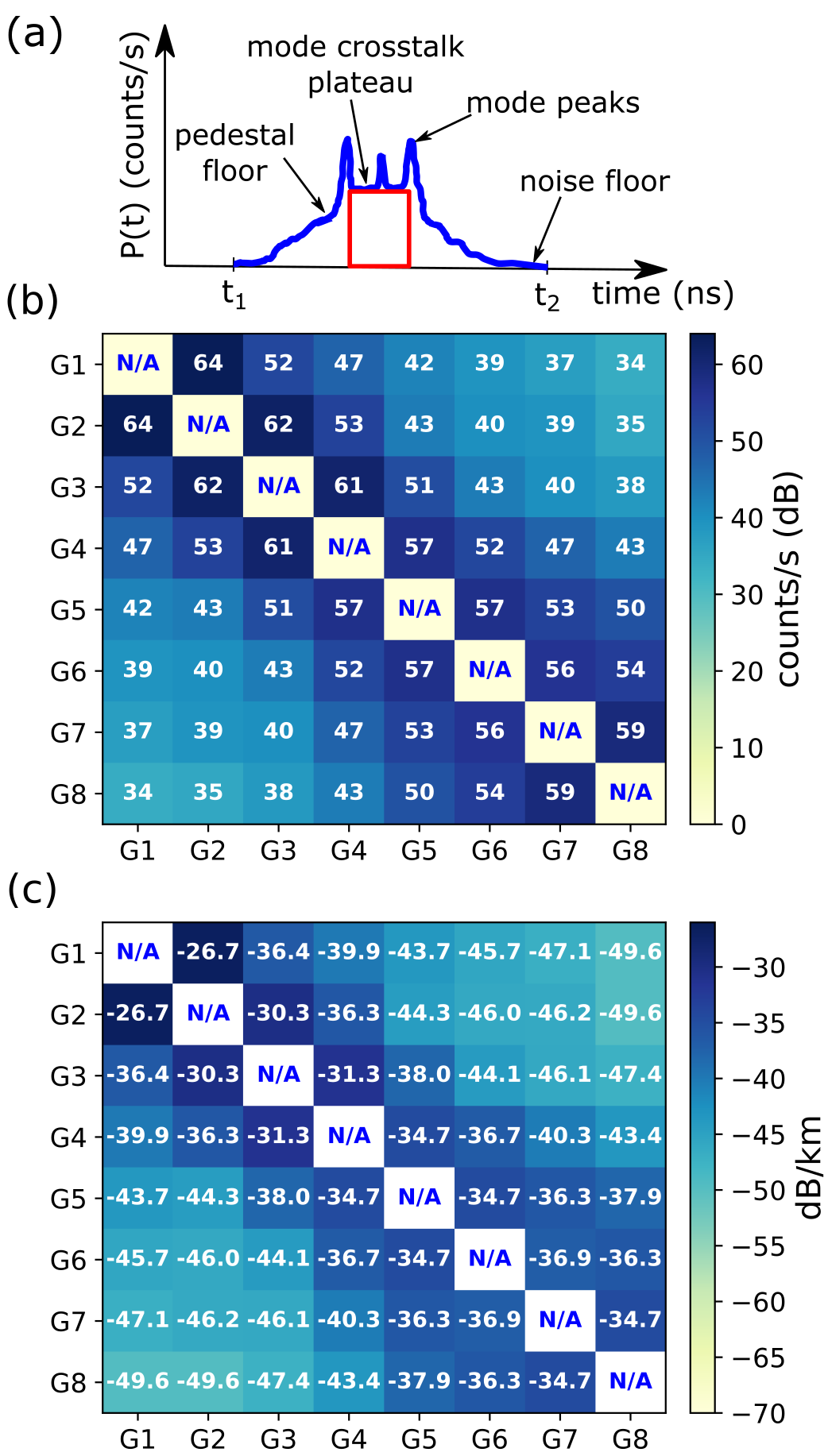}
    \caption{Coupling strength of distributed crosstalk plateau. (a) Schematic of the impulse response histogram used for normalization.The red box represents the integration area for total counts in the crosstalk plateau. (b) Distributed crosstalk plateau values averaged over all polarizations and all degenerate modes in every cell of Fig. 5 for the first 8 mode groups. (c) Distributed crosstalk plateau coefficients normalized to pulse energy and fiber length.}
    \label{fig_crosstalk_coeff}
\end{figure}

%% =====================================================

Mode coupling can occur both in the transmission fibers or in the mode (de)multiplexers. The distributed mode coupling, however, only reflects the coupling properties of the MMF. To study the distributed coupling strength in the commercial MMF, we analyze the mode crosstalk plateau between peaks in Fig. 5. The plateau values in counts/s, averaged over all polarizations and degenerate modes within each cell for the first 8 mode groups, are shown in Fig. 8(b), and the normalized distributed crosstalk coefficients are shown in Fig. 8(c). The details of this normalization are described below. 

 Fig. 8(a) schematically shows an impulse response histogram with multiple mode peaks and the distributed crosstalk plateau in between. The $y$-axis of the histogram has a unit of counts/s, corresponding to optical power. Thus the integral over time gives total photon counts, corresponding to the total energy of a specific time window. The red box in Fig. 8(a) indicates the integral area for energy in the crosstalk plateau \(E_{plateau}\). DMDG values measured from Fig. 7 are used to determine the start and end time of the plateau when calculating the integral area. If we denote the $x$-$x$ polarization curve in cell $(i,j)$ of the original 45$\times$45 matrix as \(P_{ij,xx}(t)\), the total energy of the curve is given by
\begin{equation}
E_{ij,xx}=\int_{t_1}^{t_2} P_{ij,xx}(t)dt
\end{equation}

with \(i = 1, 2, \ldots, 45, j = 1, 2, \ldots, 45\).  \( t_1\) and \(t_2\) are start and end time of the selected time window, respectively. Then we average over all four polarization curves in each cell, and sum up all columns in each row of the matrix to get pulse energy of each input channel with a specific received polarization state. Finally, we average the pulse energy of 45 input channels and use that number to normalize the distributed crosstalk plateau.
\begin{equation}
\bar E_{pulse} = \frac{1}{45}\sum_{i=1}^{45} \sum_{j=1}^{45} \frac{E_{ij,xx}+E_{ij,xy}+E_{ij,yx}+E_{ij,yy}}{4}
\end{equation}
\begin{equation}
\bar E_{plateau}(dB) = 10\times log_{10}\left(\frac{E_{plateau}}{\bar E_{pulse}}\right)
\end{equation} 

We use $t_1$ = 677 ns and $t_2$ = 717 ns, as shown in Fig. 6, which encompasses all relevant counts resulting from the pulse. The normalized distributed crosstalk coefficients are shown in Fig. 8(c). The diagonal structure of Fig. 8(b) and (c) indicates that adjacent mode groups tend to couple more strongly than groups that are far away.

% where L is the length of fiber in km. 

% Considering an interval $t_2-t_1$ = 1$\mu$s, corresponding to the period of the DML, we obtain $\bar P$ = 255647 counts/s, which is consistent with the measured count rate of $x$ and $y$ receivers at the SNSPDs. This number however, cannot be used for normalization as the pulse energy spreads less than 5\% of the entire period, making it even lower than some crosstalk values. 

%% ===================== figure 9 =======================
\begin{figure}[!h]
    \centering
    \includegraphics[width=2.7in]{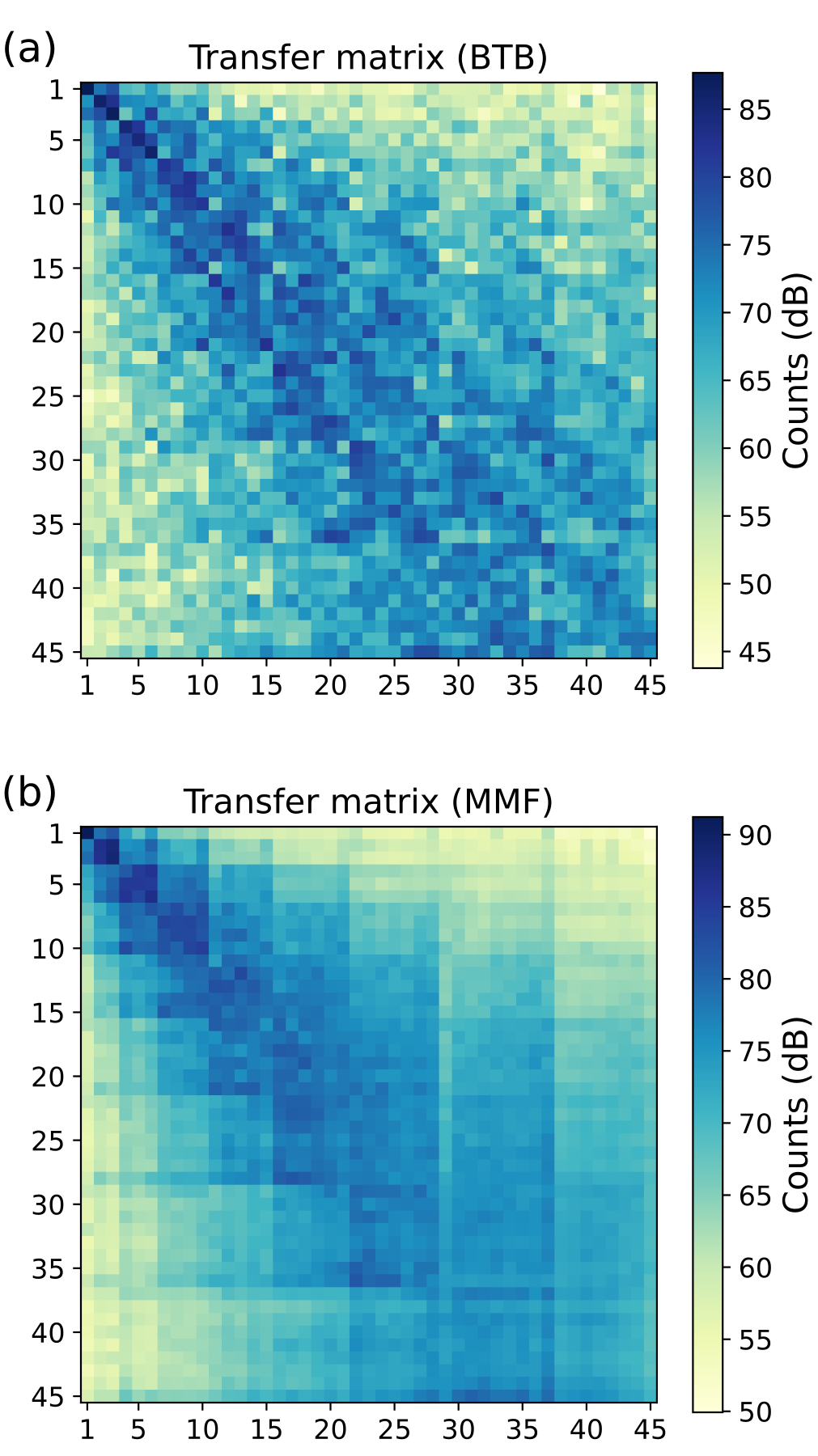}
    \caption{The complete 45$\times$45 mode transfer matrix, averaged over all polarizations of (a) BTB and (b) MMF, shows the coupling among all modes. (b) is smoother than (a) due to the mode mixing and modal dispersion in the 8.85km MMF. Note that (a) and (b) are plotted using different colormap ranges intentionally because BTB and MMF have slightly different photon count rates (input powers).}
    \label{fig_tranferMatrix}
\end{figure}

The complete 45$\times$45 mode transfer matrix of BTB and MMF, averaged over all polarizations, are shown in Fig. 9(a) and (b) respectively. For Fig. 9(a), photon count rates between 16 ns and 30 ns are integrated using Eq. (1), while for Fig. 9(b), photon count rates between 691 ns and 705 ns are integrated. The result of counts summation in this 14 ns time window represents the energy transfer matrix of the entire system, giving mode coupling information of both the MPLC mode (de)multiplexer and the MMF. Fig. 9(b) is smoother than Fig. 9(a) due to the mode mixing and modal dispersion in the 8.85km MMF. Note that the mode mixing in Fig. 9(a) is partly due to the short pigtail fiber at the output of each MPLC, and partly due to the imperfection of the packaged MPLC mode demultiplexer, as shown in Fig. 2(c).

\section{Conclusion}
In conclusion, SNSPDs offer unprecedented opportunity to study weak modal interactions in multimode optical fibers. Using SNSPD and two 45-mode MPLCs, we measure the mode transfer matrix of a commercial OM3 multimode fiber with the time-of-flight method. High sensitivity (-136 dBm) and high timing accuracy ($\sim$ 10 ps) give us unprecedented capability to accurately measure very weak modal dynamics, including distributed mode coupling, differential mode group delay, and cladding modes of multimode fibers. This research is important for applications like quantum key distribution in multimode fibers.

% or


% \appendices

% % you can choose not to have a title for an appendix
% % if you want by leaving the argument blank
% \section{}

\section*{Acknowledgment}
Y. Zhang thanks Pouria Sanjari (University of Pennsylvania, Nokia Bell Labs summer intern 2021) for help on the measurement of timing jitter of quTAG, and thanks Rene-Jean Essiambre (Nokia Bell Labs) for thoughtful discussions on the measured pedestal floor. 

\section{Supplementary}
We provide large figures of the entire 45$\times$45 mode transfer matrix of the BTB (3$\sim$43 ns) and MMF (677$\sim$ 717 ns) measurement in the \textbf{supplementary}. For BTB the 2nd peak appearing at about 33 ns, corresponding to a round trip fiber delay of 1.5 m, was due to reflections from the experimental setup \cite{mazur2021impulse}. As in the BTB case, in MMF the 2nd peak at 706 ns is also due to reflections in the system.   

% \ifCLASSOPTIONcaptionsoff
%   \newpage
% \fi



%\bibliographystyle{IEEEtran}

%% method 1: ========= use .bib file ================
\bibliographystyle{IEEEtran}
\bibliography{refs.bib}


%%%%%%%%%%%%%%%%%%%%%%%%%%%%%%%%%%%%%%%%%%%%%%%%%%%%%%%%%%%%%%%%%%%%%%%%%%%%%%%%

%%%%%%%%%%%%%%%%%%%%%%%%%%%%%%%%%%%%%%%%%%%%%%%%%%%%%%%%%%%%%%%%%%%%%%%%%%%%%%%%

%%%%%%%%%%%%%%%%%%%%%%%%%%%%%%%%%%%%%%%%%%%%%%%%%%%%%%%%%%%%%%%%%%%%%%%%%

\end{document}